\documentclass[aps,prx,reprint]{revtex4-2}
\usepackage{amsmath}
\usepackage{amsfonts}
\usepackage{mathtools}

\usepackage[colorlinks=true, pdfstartview=FitV, linkcolor=blue,
citecolor=blue, urlcolor=blue]{hyperref}
\usepackage[capitalise]{cleveref}

\usepackage{amssymb}
\usepackage{siunitx}
\usepackage{physics} 
\usepackage{graphicx}
\usepackage{xcolor}

\newcommand{\vect}[1]{\boldsymbol{#1}}

\begin{document}

	\title{Microscopic analysis of proximity-induced superconductivity and metallization effects in superconductor-germanium hole nanowires}
	
	\author{Christoph Adelsberger}
	\author{Henry F. Legg}
	\author{Daniel Loss}
	\author{Jelena Klinovaja}
	\affiliation{Department of Physics, University of Basel, Klingelbergstrasse 82, CH-4056 Basel, Switzerland}

\begin{abstract}
Low-dimensional germanium hole devices are promising systems with many potential applications such as hole spin qubits, Andreev spin qubits, and Josephson junctions, and can serve as a basis for the realization of topological superconductivity.  This vast array of potential uses for Ge largely stems from the exceptionally strong and controllable spin-orbit interaction (SOI), ultralong mean free paths, long coherence times, and compatibility with complementary metal-oxide-semiconductor (CMOS) technology. However, when brought into proximity with a superconductor (SC), metallization normally diminishes many useful properties of a semiconductor, for instance, typically reducing the $g$ factor and SOI energy, as well as renormalizing the effective mass. In this paper, we consider metallization of a Ge nanowire (NW) in proximity to a SC, explicitly taking into account the three-dimensional (3D) geometry of the NW. We find that proximitized Ge exhibits a unique phenomenology of metallization effects, where the 3D cross section plays a crucial role. For instance, in contrast to expectations, we find that SOI can be enhanced by strong coupling to the superconductor. We also show that  the thickness of the NW plays a critical role in determining both the size of the proximity-induced pairing potential and metallization effects, since the coupling between the NW and SC strongly depends on the distance of the NW wave function from the interface with the SC. In the absence of electrostatic effects, we find that a sizable gap opens only in thin NWs ($d\lesssim \SI{3}{\nano\meter}$). In thicker NWs, the wave function must be pushed closer to the SC by electrostatic effects in order to achieve a sizable proximity gap such that the required electrostatic field strength can simultaneously induce a strong SOI. The unique and sometimes beneficial nature of metallization effects in SC-Ge NW devices evinces them as ideal platforms for future applications in quantum information processing.
\end{abstract}
	
	\maketitle

\section{Introduction}
Hole gases in Ge heterostructures are one of the most promising platforms for applications in quantum information processing~\cite{Scappucci2021}. The compatibility of Ge with Si complementary metal-oxide-semiconductor (CMOS) technology is important to achieve the scalability that is required for the building blocks of any future quantum computer~\cite{Veldhorst2017}. In addition, however, holes in Ge have many favorable properties, such as the possibility to grow ultraclean substrates with a mean free path up to $\SI{30}{\micro\meter}$~\cite{Myronov2023,Stehouwer2023,Sammak2019,Kamata2008}, the long spin coherence times due to weak hyperfine noise suppressed by appropriate quantum dot design~\cite{Fischer2010,Maier2012,Klauser2006,Prechtel2016,Testelin2009,Fischer2008,Bosco2021a} or isotopic purification~\cite{Itoh1993, Itoh2014}, the strong and tunable spin-orbit interaction (SOI)~\cite{Kloeffel2018,Kloeffel2011,Hao2010,Hu2011,Terrazos2021,Froning2021a,Liu2022}, and the tunable $g$ factor~\cite{Maier2013,Bosco2022a,Adelsberger2022,Adelsberger2022a}. These properties have already enabled the realization of high-quality spin qubits~\cite{Loss1998} with which electrically controlled single-~\cite{Wang2022,Watzinger2018,Hendrickx2020a,Froning2021} and two-qubit~\cite{Hendrickx2020} gates, as well as singlet-triplet encoding~\cite{Jirovec2021} and a four-qubit processor~\cite{Hendrickx2021}, have been demonstrated.

\begin{figure}[]
	\includegraphics[width=\columnwidth]{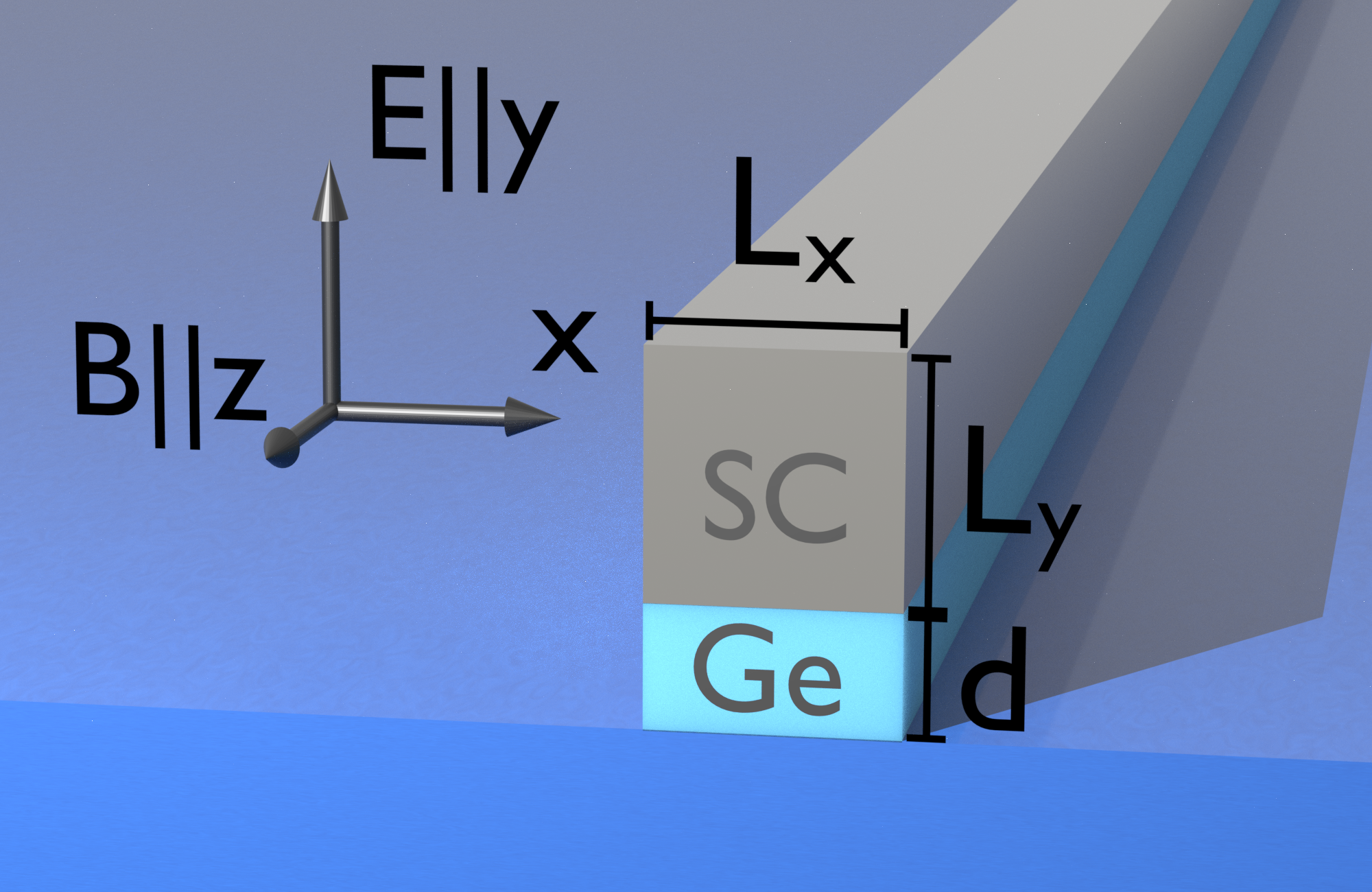}
	\caption{Sketch of a rectangular Ge NW in proximity to a superconductor (SC). The SC cross section is defined by $L_x = L_y = \SI{10}{\nano\meter}$ and the Ge NW cross section by the width $L_x= \SI{10}{\nano\meter}$ and height $d$, which is not fixed and will be used as a parameter in what follows. We assume an infinitely long system in the $z$ direction.  A magnetic field $B$ is applied along the NW axis in the $z$ direction, and an electric field $E$ is applied in the perpendicular $y$ direction. \label{fig:schematic}}
\end{figure}

In the last few years, there has been considerable progress in fabricating hybrid devices that couple Ge to superconductors (SCs) such as aluminum~\cite{Xiang2006,Hendrickx2018,Ridderbos2018,Hendrickx2019,Vigneau2019,Ridderbos2019,Aggarwal2021,Tosato2023,Zhuo2023} (see Fig.~\ref{fig:schematic}). Hybrid SC-Ge devices substantially increase the possible applications of Ge, for instance, enabling electrically controllable Josephson junctions~\cite{Xiang2006,Ridderbos2019a,Casparis2018,Larsen2020}, allowing for long-range coupling of spin qubits~\cite{Petersson2012,Leijnse2013,Larsen2015,Burkard2020}, providing the basis for Andreev spin qubits~\cite{Lee2013,Park2017,Hays2021,Spethmann2022}, and as a platform to realize topological superconductivity with associated Majorana bound states (MBSs)~\cite{Kitaev2001,Ivanov2001,Nayak2008,Maier2014,Pientka2017,Lutchyn2018,Luethi2023,Luethi2023a,Woods2023,Laubscher2023}.

The coupling of a superconductor to a semiconductor, however, not only results in a proximity-induced superconducting pairing potential~\cite{Cole2015,Stanescu2017} but has additional consequences due to metallization of the semiconductor by the SC~\cite{Reeg2018}. Such effects, e.g., the renormalization of the effective mass, the $g$ factor, and the spin-orbit energy, have been widely studied  in platforms expected to achieve MBSs~\cite{Reeg2017,Reeg2017a,Reeg2018,Antipov2018,Mikkelsen2018,Moor2018,Woods2018,Winkler2019,Awoga2019,Vaitiekenas2020,Aghaee2023,Legg2022}.
The most extensively investigated platform to achieve topological superconductivity is semiconducting nanowires (NWs) with strong Rashba SOI, such as InAs and InSb~\cite{Laubscher2021}. In such systems, it has been predicted that the opening of a proximity-induced gap in the NW also results in a reduction of the SOI energy and the $g$ factor~\cite{Reeg2017, Reeg2018}.  When the SC strongly couples to the semiconductor, metallization effects can make it difficult to find a regime capable of hosting MBSs because this simultaneously requires strong SOI, a large Zeeman energy, and a sizable proximity-induced pairing potential. Due to the unique phenomenology of holes in Ge, it was not yet clear what the consequences of metallization effects are in Ge and what limitations metallization places on the potential applications of Ge based platforms.

In this paper, we numerically investigate the metallization of holes in three-dimensional (3D) Ge NWs that have been brought into proximity with a 3D SC. Importantly, the wave function is nonuniformly distributed throughout the NW cross section. As a result, we find the thickness of the Ge NW plays a crucial role in both the size of the induced proximity gap and the consequences of metallization effects. In the absence of electrostatic fields, we show that only when the NW is very flat ($d\lesssim\SI{3}{\nano\meter}$), such that the wave functions of states in the NW are close to the SC, can a sizable proximity-induced gap be achieved. However, for thicker NWs, electrostatic fields, e.g., due to gating or interface effects, can push the wave functions close to the Ge-SC interface, increasing the proximity-induced gap and with important separate consequences for the SOI strength. In particular, we will demonstrate that it is possible to enhance the SOI and the proximity-induced gap at the same time by tuning with an external electric field. Furthermore, using thicker NWs has the advantage that one can reach the optimal side length ratio for maximal SOI that depends on the electric field~\cite{Bosco2021,Adelsberger2022} and that the $g$ factor is larger than in flat NWs. The mechanism behind the strong SOI is the direct Rashba SOI typically present in one-dimensional (1D) hole systems~\cite{Kloeffel2011,Kloeffel2018,Adelsberger2022,Adelsberger2022a}. Our results show that metallization effects in SC-Ge hole hybrid devices have a unique phenomenology and are often either benign or even beneficial in nature. Our findings suggest that although metallization effects can have important consequences for SC-Ge devices, such hybrid systems remain a promising avenue for future quantum information processing applications. 

The focus of this work lies in analyzing effective models for proximity-induced superconductivity and metallization in Ge hole NWs. These models facilitate the understanding of the qualitative behavior of the parameters and their mutual dependences, including the proximity-induced superconducting order parameter, the NW $g$ factor, and the strength of the SOI. However, certain factors lie outside the scope of our study, such as the electrostatic potential resulting from band bending at the interface between the SC and the semiconductor, charges originating from impurities, or disorder. A comprehensive analysis of additional electrostatic effects, e.g., due to precise work function differences of atoms at the interface, necessitates computationally intensive \textit{ab-initio} calculations such as density functional theory~\cite{Ruessmann2022}, which exceeds the boundaries of our current work. Although all the aforementioned phenomena are anticipated to lead to quantitative adjustments in our results, the qualitative behavior that we present here is expected to remain unchanged. In particular, the band bending at the interface, typically addressed in numerical Schr\"{o}dinger-Poisson calculations~\cite{Antipov2018,Winkler2019,Mikkelsen2018,Woods2018}, gives rise to an accumulation or repulsion of charges at the interface. Accounting for these effects is expected to modify the strength of the external electric field required to confine the hole wave function at the interface, as discussed in Sec.~\ref{sec:elecField}; the outcome, however, would remain the same.

This paper is structured as follows. In Sec.~\ref{sec:model}, we introduce our theoretical model of a Ge hole NW coupled to a SC. In Sec.~\ref{sec:SCchempot}, we analyze the influence of the NW-SC coupling on the Fermi wave vector of the NW states. We show that the chemical potential of the SC with respect to the chemical potential of the NW decides whether the NW Fermi wave vector is increased or decreased with increasing coupling. We decide for the case where the Fermi wave vector of the NW is decreased with increasing coupling as the most relevant one. In Sec.~\ref{sec:NWthick}, we investigate the consequences of changing the thickness of the NW. Our results show that without electrostatic fields, only thin NWs couple strongly to the SC, which can be explained by the distance of the wave function in the NW from the SC-Ge interface. In thick NWs, however, the wave function can be localized close to the SC by an external electric field, as we show in Sec.~\ref{sec:elecField}.  As a consequence, the spin-orbit energy and the induced gap can simultaneously increase with increasing electric field. A discussion of the coupling mechanism between NW and SC is given in Sec.~\ref{sec:noHH}. Finally, in Sec.~\ref{sec:conclusion}, we conclude and give an outlook of implications for Ge-based superconducting devices. 
	
\section{Model \label{sec:model}}
In this section, we introduce the model that we employ to describe a Ge NW that is coupled to a SC placed on top of the NW, with both NW and SC extending infinitely along the $z$ direction (see Fig.~\ref{fig:schematic}). We model the cross section of the system by a finite 2D lattice in real space. The momentum $\hbar k$ along the $z$ axis is a good quantum number since we assume translational invariance in this direction and periodic boundary conditions.  The discrete model for the NW coupled to a SC is then
\begin{align}
	H(k) = H_w(k) + H_s(k) + H_c(k),
\end{align}
where $ H_w(k)$ is the Hamiltonian for the hole NW, $H_s(k)$ is the Hamiltonian for the SC, and $H_c(k)$ is the Hamiltonian describing the tunnel coupling between the NW and the SC at the interface.

To describe the heavy hole (HH) and light hole (LH) nature of the Ge NW, we use the isotropic Luttinger-Kohn Hamiltonian,
\begin{align}
	H_\mathrm{LK} = -\frac{\hbar^2}{2 m_e} \left[\gamma_k k^2 - 2 \gamma_s (\vect{k} \cdot \vect{J})^2\right], \label{eqn:LK_Hamiltonian_sphericalApp}
\end{align}
as commonly utilized in the literature to describe the states in Ge~\cite{Kloeffel2018,Kloeffel2011,Bosco2021,Lipari1970,Li2021,Li2021b};
here, $\gamma_k = \gamma_1+5\gamma_s/2$, $\gamma_s = (\gamma_2 + \gamma_3)/2 = 4.97$, and $J_i$ [with $i = x, y, z$] are the standard spin-$3/2$ operators. In reality, the holes in Ge are not spin-$3/2$ particles, but their total angular momentum is $j = l + s = 3/2$, where $l$ is the orbital angular momentum that is $l=1$ for a $p$-type orbital and $s=1/2$ is the spin.  The coefficients $\gamma_1=13.35$, $\gamma_2=4.25$, and $\gamma_3=5.69$ are the material-dependent Luttinger parameters~\cite{Winkler2003} and $m_e$ is the free electron mass. Note the global negative sign in Eq.~\eqref{eqn:LK_Hamiltonian_sphericalApp} for holes. 

To lift the Kramers degeneracy, we add a small Zeeman field in the $z$ direction that enters via the  Zeeman Hamiltonian
\begin{align}
	H_Z = 2 \kappa \mu_B B J_z,\label{eqn:HamZeeman}
\end{align}
with magnetic field strength $B$ and $\kappa=3.41$ in Ge~\cite{Lawaetz1971}. Here we neglect the small anisotropic Zeeman energy $\propto J_i^3$~\cite{Luttinger1956, Kloeffel2018} and effects of orbital magnetic fields~\cite{Lim2012,Osca2015,Nijholt2016,Dmytruk2018,Wojcik2018} since we consider only very weak magnetic fields applied parallel to the NW. Furthermore, we include a homogeneous electric field in the $y$ direction via the Hamiltonian
\begin{align}
	H_E = -e E y,
\end{align}
where $E$ is the strength of the electric field in the $y$ direction.

On the 2D square lattice that models the cross section of the NW, the NW Hamiltonian becomes 
\begin{align}
	H_{w, k} = &-\sum_{n=1, m=1}^{L_x/a,d/a} \vect{c}_{n,m,k}^\dagger \Big[ H_\mathrm{LK}^{k_z^2} k^2 + \frac{2}{a^2} \left(H_\mathrm{LK}^{k_x^2}+ H_\mathrm{LK}^{k_y^2}\right) \nonumber\\
	&+ H_Z +e E a m - \mu_w\Big] \vect{c}_{n,m,k}\nonumber\\
	& +\Big[ \vect{c}_{n+1,m,k}^\dagger \left(\frac{i}{2 a} H_\mathrm{LK}^{k_xk_z} k -\frac{1}{a^2} H_\mathrm{LK}^{k_x^2}\right)\vect{c}_{n,m,k}  \nonumber\\
	& + \vect{c}_{n,m+1,k}^\dagger \left(\frac{i}{2 a} H_\mathrm{LK}^{k_yk_z} k -\frac{1}{a^2} H_\mathrm{LK}^{k_y^2}\right)\vect{c}_{n,m,k} \nonumber\\
	&- \vect{c}_{n+1,m+1,k}^\dagger \frac{1}{4 a^2} H_\mathrm{LK}^{k_xk_y} \vect{c}_{n,m,k} + \mathrm{H.c.}\Big],
\end{align}
with
\begin{align}
	H_\mathrm{LK}^{k_x^2} &= \frac{\hbar^2}{m_e}\begin{pmatrix}
		\frac{\gamma_1+\gamma_s}{2} & 0 & -\frac{\sqrt{3} \gamma_s}{2} & 0 \\
		0 & \frac{\gamma_1-\gamma_s}{2}  & 0 & -\frac{\sqrt{3} \gamma_s}{2}\\
		-\frac{\sqrt{3} \gamma_s}{2} & 0 & \frac{\gamma_1-\gamma_s}{2}  & 0\\
		0 & -\frac{\sqrt{3} \gamma_s}{2} & 0 & \frac{\gamma_1+\gamma_s}{2}
	\end{pmatrix}, \label{eqn:LKHam_kx2}\\
H_\mathrm{LK}^{k_y^2} &= \frac{\hbar^2}{m_e}\begin{pmatrix}
	\frac{\gamma_1+\gamma_s}{2} & 0 & \frac{\sqrt{3} \gamma_s}{2} & 0 \\
	0 & \frac{\gamma_1-\gamma_s}{2}  & 0 & \frac{\sqrt{3} \gamma_s}{2}\\
	\frac{\sqrt{3} \gamma_s}{2} & 0 & \frac{\gamma_1-\gamma_s}{2}  & 0\\
	0 & \frac{\sqrt{3} \gamma_s}{2} & 0 & \frac{\gamma_1+\gamma_s}{2}
\end{pmatrix},\label{eqn:LKHam_ky2}
\end{align}
\begin{align}
H_\mathrm{LK}^{k_z^2} &= \frac{\hbar^2}{m_e}\begin{pmatrix}
	\frac{\gamma_1-2\gamma_s}{2} & 0 & 0 & 0 \\
	0 & \frac{\gamma_1+2\gamma_s}{2}  & 0 & 0\\
	0 & 0 & \frac{\gamma_1+2\gamma_s}{2}  & 0\\
	0 & 0 & 0 & \frac{\gamma_1-2\gamma_s}{2}
\end{pmatrix},\label{eqn:LKHam_kz2}\\
H_\mathrm{LK}^{k_xk_y} &= \frac{\hbar^2}{m_e}\begin{pmatrix}
	0& 0 & i\sqrt{3} \gamma_s & 0 \\
	0 & 0  & 0 & i\sqrt{3} \gamma_s\\
	-i\sqrt{3} \gamma_s & 0 & 0 & 0\\
	0 & -i\sqrt{3} \gamma_s & 0 & 0
\end{pmatrix},\label{eqn:LKHam_kxky}\\
H_\mathrm{LK}^{k_xk_z} &= \frac{\hbar^2}{m_e}\begin{pmatrix}
	0& -\sqrt{3} \gamma_s & 0 & 0 \\
	-\sqrt{3} \gamma_s & 0  & 0 &0\\
	0 & 0 & 0 & \sqrt{3} \gamma_s\\
	0 &0& \sqrt{3} \gamma_s & 0
\end{pmatrix},\label{eqn:LKHam_kxkz}\\
H_\mathrm{LK}^{k_yk_z} &= \frac{\hbar^2}{m_e}\begin{pmatrix}
	0& i\sqrt{3} \gamma_s & 0 & 0 \\
	-i\sqrt{3} \gamma_s & 0  & 0 &0\\
	0 & 0 & 0 & -i \sqrt{3} \gamma_s\\
	0 &0& i\sqrt{3} \gamma_s & 0
\end{pmatrix}.\label{eqn:LKHam_kykz}
\end{align}
We define the four-dimensional vectors $\vect{c}_{n,m,k}^\dagger = \left(c_{+\frac{3}{2}}^\dagger,c_{+\frac{1}{2}}^\dagger,c_{-\frac{1}{2}}^\dagger,c_{-\frac{3}{2}}^\dagger\right)_{n,m,k}$, where $c_{\pm\frac{3}{2} (\frac{1}{2})}^\dagger$ describes the creation of a hole with total $J_z$ angular momentum $\pm \frac{3}{2} (\frac{1}{2})$.
The sum runs over all sites of the lattice $(n,m)$ where the indices $n$ and $m$ run over the $x$ and $y$ coordinates, respectively. We measure the chemical potential $\mu_w$ from the Rashba crossing point at $k=0$. For the calculations we choose the lattice spacing $a= \SI{0.1}{\nano\meter}$. 

The discretized Hamiltonian for the conventional $s$-wave superconductor is given by
\begin{align}
	H_{s,k} = &\sum_{n=1, m=1,\sigma=\uparrow, \downarrow}^{L_x/a, L_y/a} b_{n,m,k,\sigma}^\dagger \bigg[ \frac{\hbar^2}{2 m_s}\left(k^2 + \frac{2}{a^2}\right)\nonumber\\
	 &+ \frac{g_s \mu_B}{2} B \sigma_z- \mu_s\bigg] b_{n,m,k,\sigma}\nonumber\\
	& - \frac{\hbar^2}{2 m_s}\Bigg(\sum_{\substack{m=1, \sigma=\uparrow, \downarrow\\ \left<n', n\right>}}^{L_y/a} b_{n', m,k, \sigma}^\dagger  \frac{1}{a^2} b_{n, m, k,\sigma}\nonumber\\
	 &+ \sum_{\substack{n=1, \sigma=\uparrow, \downarrow\\ \left<m', m \right>}}^{L_x/a} b_{n, m',k, \sigma}^\dagger  \frac{1}{a^2} b_{n, m,k, \sigma}\Bigg)\nonumber\\
	 &+\sum_{n=1,m=1}^{L_x/a, L_y/a} \Big(\Delta_0 b_{n,m,k,\uparrow}^\dagger b_{n,m,-k,\downarrow}^\dagger \nonumber\\
	 &+  \Delta_0^\ast b_{n,m,-k,\downarrow} b_{n,m,k,\uparrow} \Big),\label{eq:SC}
\end{align}
where $b_{n,m,k,\sigma}^\dagger (b_{n,m,k,\sigma})$ creates an electron (hole) with spin $\sigma= \uparrow, \downarrow$ in the superconductor. The effective mass of the superconductor is $m_s$ and, in addition, we take the superconducting pairing potential as $\Delta_0 = \SI{0.2}{\milli\electronvolt}$. The expression $\left<n', n\right> $ $(\left<m', m \right>)$ describes a sum over neighboring sites in the $x (y)$ direction. We measure the chemical potential of the SC, $\mu_s$, from the bottom of the lowest subband and we choose the effective mass $m_s = 0.95 m_e$. This results in the hopping amplitude $t_s = \frac{\hbar^2}{2 m_s a^2} \approx \SI{4}{\electronvolt}$ and Fermi velocity $v_{F,s} = \partial_k \mathcal{E}(k)|_{k=k_F}/\hbar \approx \SI{1.27E6}{\meter\per\second}$ for  $\mu_s = \SI{8.75}{\electronvolt}$, where 
$\mathcal{E}(k) = (t_s/a^2) [1-\cos(k a) ]- \mu_s$ is the dispersion relation.

\begin{figure*}[]
	\includegraphics[width=\textwidth]{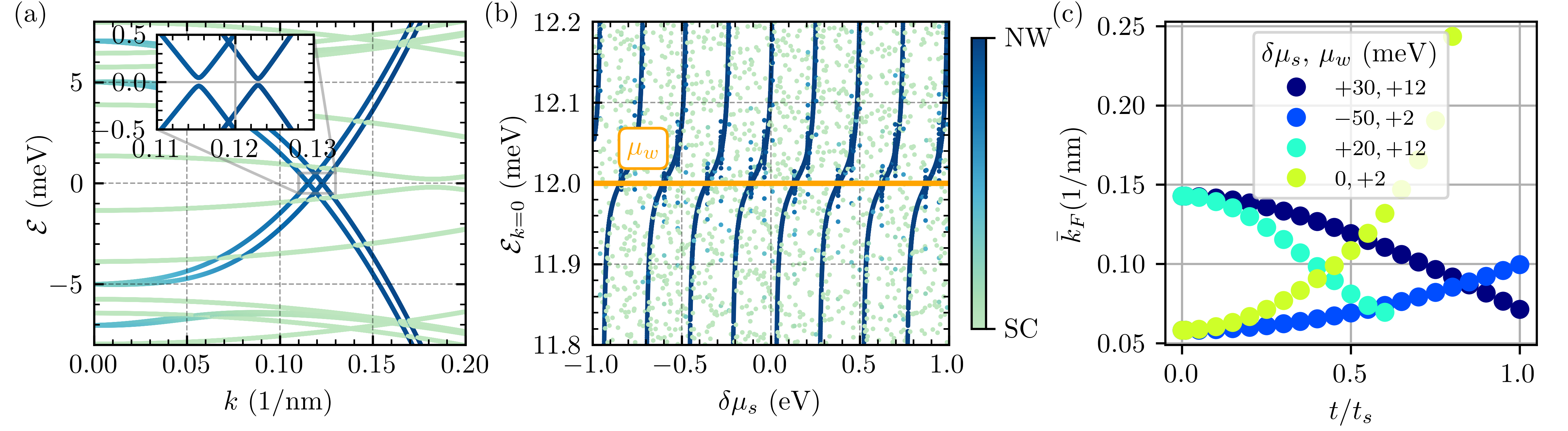}
	\caption{(a) Dispersion relation of the SC-Ge hole NW for $t= 0.5 t_s$, where the color bar to the right of (b) shows the weight of the wave function in the NW and SC for (a) and (b).  The gap opening in the NW band is shown in the inset in (a). Due to finite SOI for nonzero SC-NW coupling $t$, there are two Fermi momenta. The chemical potential of the SC is fixed to $\mu_s = \SI{8.77}{\electronvolt}$. In order to achieve coupling between the NW  and SC states, they have to fulfill certain selection rules given by the confinement and the related quantum numbers~\cite{Reeg2017,Reeg2018}. This is the reason why most of the SC subband states cross the NW subbands without hybridization, so without gap opening.
		(b)~Energies of the SC-Ge hole NW states at $k=0$ as a function of the shift of the chemical potential of the SC, $\delta \mu_s$, measured from $\mu_s = \SI{8.75}{\electronvolt}$ for $t=0.1 t_s$. Here, we set $\Delta_0=0$ and consider a metal instead of a SC. The energies of the NW states depend strongly on $\delta \mu_s$. For roughly half of the $\delta \mu_s$ values, the energy of the NW states at $k=0$ increases compared to the chemical potential of the uncoupled NW, $\mu_w = \SI{12}{\milli\electronvolt}$ (indicated by the orange line). For the other half of the $\delta \mu_s$ values, the energy of the NW states at $k=0$ decreases.
(c) As a consequence of the observed energy behavior, the average Fermi wave vector $\bar{k}_F$ can either increase or decrease with $t$, depending on the value of $\delta \mu_s$. For instance, the decrease in energy for $\delta \mu_s = \SI{+30}{\milli\electronvolt}$ (dark blue) causes the initial Fermi wave vector to decrease with increasing coupling $t$ to the SC. For $\delta \mu_s = \SI{-50}{\milli\electronvolt}$ (light blue), the average Fermi wave vector increases with growing coupling to the SC. For $\delta \mu_s = 0$ (green) and $\delta \mu_s = \SI{+20}{\milli\electronvolt}$ (cyan), the NW state at $k=0$ is very close in energy to a SC state that couples to the NW. At this resonance, $\bar{k}_F$ changes very strongly with $t$. If not stated differently, the parameters used are $\mu_w = \SI{12}{\milli\electronvolt}$, $\mu_s = \SI{8.75}{\electronvolt}$, $d = \SI{2}{\nano\meter}$, $L_x=L_y =\SI{10}{\nano\meter}$, $a= \SI{0.1}{\nano\meter}$,  $\Delta_0=\SI{0.2}{\milli\electronvolt}$, $E=0$, and $B=\SI{0.01}{\tesla}$. \label{fig:dispRelmuSC}}
\end{figure*}

Since the SC Hamiltonian is given in a spin basis and the NW Hamiltonian in a total angular momentum basis, we cannot couple them without first  applying a basis transformation. In the following, for simplicity of presentation, we utilize the Hamiltonian $H_M$ of a simple metal that is the same as Eq.~\eqref{eq:SC} with no pairing potential, $\Delta_0=0$. The total $J_z$ angular momentum basis states $P^{-1}\ket{\psi}=(\ket{+3/2}, \ket{+1/2}, \ket{-1/2}, \ket{-3/2})$ of the NW are given in terms of the orbital angular momentum and spin $\ket{\psi} = \ket{l_z, s_z}$ by~\cite{Yu2010}
\begin{align}
	\ket{+3/2} &= \ket{+1, \uparrow},\nonumber\\
	\ket{+1/2} &= \frac{1}{\sqrt{3}}\left(\ket{+1, \downarrow} + \sqrt{2}\ket{0, \uparrow}\right),\nonumber\\
	\ket{-1/2} &= \frac{1}{\sqrt{3}} \left(\ket{-1, \uparrow} + \sqrt{2} \ket{0, \downarrow}\right), \nonumber\\
	\ket{-3/2} &= \ket{-1, \downarrow}. \label{eqn:LKHamBasis}
\end{align}
Note that the two spin-orbit split-off states ($J=1/2$) are neglected here.
The unitary matrix
\begin{align}
P = P^{-1}= \begin{pmatrix}
		1 & 0 & 0 & 0 \\
		0 & \frac{1}{\sqrt{2}} & \frac{1}{\sqrt{2}} & 0 \\
		0 & \frac{1}{\sqrt{2}} & -\frac{1}{\sqrt{2}} & 0 \\
		0 & 0 & 0 & 1
	\end{pmatrix}
\end{align}
transforms the total angular momentum basis such that we can write the coupling between one site of the metal $H_{M,i}$ to one site of the NW $H_{w,i}$ as the matrix
\begin{align}
	\begin{pmatrix}
		P^{-1} H_{w,i} P & H_t \\
		H_t^T & H_{M,i}
	\end{pmatrix}, \label{eqn:NWMcoupHam}
\end{align}
where the basis of this matrix is $(\ket{\psi}, \ket{\uparrow}, \ket{\downarrow})$ and the coupling matrix is 
\begin{align}
	H_t = \begin{pmatrix}
		t_\mathrm{HH} & t_\mathrm{LH} & 0 & 0\\
		0 & 0 & t_\mathrm{LH} & t_\mathrm{HH}
	\end{pmatrix}^T, \label{eqn:NWSCcoup}
\end{align}
with the LH (HH) coupling amplitudes $t_\mathrm{LH}$ ($t_\mathrm{HH}$).  This is a simplified model for the coupling between a semiconductor NW and a metal, but sufficient to capture the qualitative physics of metallization effects in Ge. We further simplify this by assuming HH and LH coupling amplitudes, i.e., $t= t_\mathrm{HH} = t_\mathrm{LH}$. For an analysis of the situation where $t_\mathrm{LH}\neq 0 $ and $t_\mathrm{HH}=0$, see Sec.~\ref{sec:noHH}. In  general, the coupling amplitudes are different and depend on $k$~\cite{Futterer2011,Moghaddam2014}, which is neglected here. Furthermore, in an experiment, the Ge NW would be covered by a shell that induces strain into the NW and changes the tunnel barrier between SC and Ge. Another possible realization would be a gate-defined 1D channel in a planar Ge/SiGe heterostructure~\cite{Adelsberger2022a}. However, we expect only quantitative changes of our results due to these details.

Note that the coupling $t$ between the SC and the Ge NW is a phenomenological parameter in our model, which is not an experimental observable. However, in the following we present the proximity-induced superconducting order parameter and the NW $g$ factor and SOI as a function of $t$. In an experiment the proximity-induced gap can be measured which then relates to a certain value of our model parameter $t$. This then allows us to predict the $g$ factor and SOI for the measured superconducting gap size.

\section{Chemical potential of the superconductor \label{sec:SCchempot}}

Before we discuss proximity-induced superconductivity in Ge NWs, we first analyze the dependence of the average Fermi wave vector in the NW, $\bar{k}_F = (k_F^1 + k_F^2)/2$, on the chemical potential of the SC, $\mu_s$, where $k_F^{1,2}$ are the two Fermi momenta in the presence of SOI [see Fig.~\ref{fig:dispRelmuSC}(a)]. In the uncoupled case, we can connect the Fermi wave vector to the charge carrier density $n = e(k_F^1 + k_F^2)/\pi$ in the NW with the positive elementary charge $e$ for holes. In general, the sizes of the induced gaps at the two different Fermi wave vectors are not equal. In the following, we refer to the Fermi wave vector $k_F$ denoted, without further index, as the one at which the gap is smaller. In the following, we study the shift of the energy of the state at $k=0$ and $\bar{k}_F$ of the NW on the chemical potential of the SC. However, in reality, it is challenging to control the chemical potential of the SC, but the thickness of the SC is under control. The same resonances as shown in Fig.~\ref{fig:dispRelmuSC}(b) can be observed as a function of the thickness of the SC~\cite{Reeg2018}, which is obvious because the level spacing of the SC depends on the chemical potential via the Fermi velocity as well as on the thickness of the SC.

For Fig.~\ref{fig:dispRelmuSC}(b), we set $\Delta_0=0$ and assume that the NW is coupled to a normal metal. We fix the chemical potential of the NW to $\mu_w = \SI{12}{\meV}$ and sweep the chemical potential of the metal around $\mu_s =\SI{8.75}{\electronvolt}$. In Fig.~\ref{fig:dispRelmuSC}(b), we show the energies of the NW and the metal states at $k=0$. The energies of the NW states (dark-blue dots) show resonances at certain values of $\delta \mu_s$ every time the lowest NW state at $k=0$ couples to a metal state that lies at the same energy. Note that only certain metal states couple to the NW because they need to fulfill selection rules that are given by the quantum numbers related to the confinement~\cite{Reeg2017,Reeg2018}. The periodicity of the resonances is set by the level spacing of the metal, $\pi\hbar v_{F,s}/L_y \approx \SI{263}{\milli\electronvolt}$.

From previous investigations of metallization effects in semiconductors \cite{Reeg2017, Reeg2018}, we expect that coupling to the SC causes an increase of the Fermi wave vector of the NW states. This behavior is, indeed, found in a large parameter regime of our numerical study [see Fig.~\ref{fig:dispRelmuSC}(b)]. However, within our model, shifting the chemical potential of the SC, $\mu_s$, we find cases where $\bar{k}_F$ of the NW decreases. Whether $\bar{k}_F$ increases or decreases depends on whether $\delta \mu_s$ lies on the left or the right side of a resonance in Fig.~\ref{fig:dispRelmuSC}(b). We show $\bar{k}_F$ as a function of the coupling $t$ between SC and NW for different values of $\delta\mu_s$ in Fig.~\ref{fig:dispRelmuSC}(c) to illustrate this behavior. The chemical potential of the NW, $\mu_w$, takes two different values for the curves in Fig.~\ref{fig:dispRelmuSC}(c). However, this is not the reason for the different behaviors. In the case where $\bar{k}_F$ decreases with increasing $t$, we need to set $\mu_w$ to a larger value at $t=0$ to avoid a rapid depletion of the NW. For increasing $\bar{k}_F$, this is not necessary and we start at $t=0$ with a smaller value for $\mu_w$. The cyan line ends at $t=0.6$, because at stronger coupling, the Fermi wave vector is no longer well defined as the energy of the state at $k=0$ becomes comparable to the size of the superconducting gap.

In the following, we will focus on the situation where the average Fermi wave vector $\bar{k}_F$ decreases with increasing coupling to the SC. However, the coupling itself also causes a shift of chemical potentials in the NW and the SC and thus, as we will see below, a stronger SC-NW hybridization can occur as the coupling is increased. Furthermore, we choose $\mu_s$ such that it is away from the resonances. Calculations closer to a resonance show qualitatively the same, but quantitatively stronger effects. In order to hit such a resonance in an experiment, fine tuning of the SC thickness would be required, which is difficult in practice (but not impossible, e.g., in the case of epitaxial growth of the SC on top of the NW). For the situation where $\bar{k}_F$ increases with $t$, we observe the same effects, which is not shown here. The only noteworthy difference is that the proximity-induced superconducting order parameter does not converge to a constant value as quickly as in the case of  decreasing $\bar{k}_F$. In fact, a convergence sets in only for $t>t_s$ which, however, is an unrealistic regime.

\section{Nanowire thickness \label{sec:NWthick}}

 \begin{figure*}[]
	\includegraphics{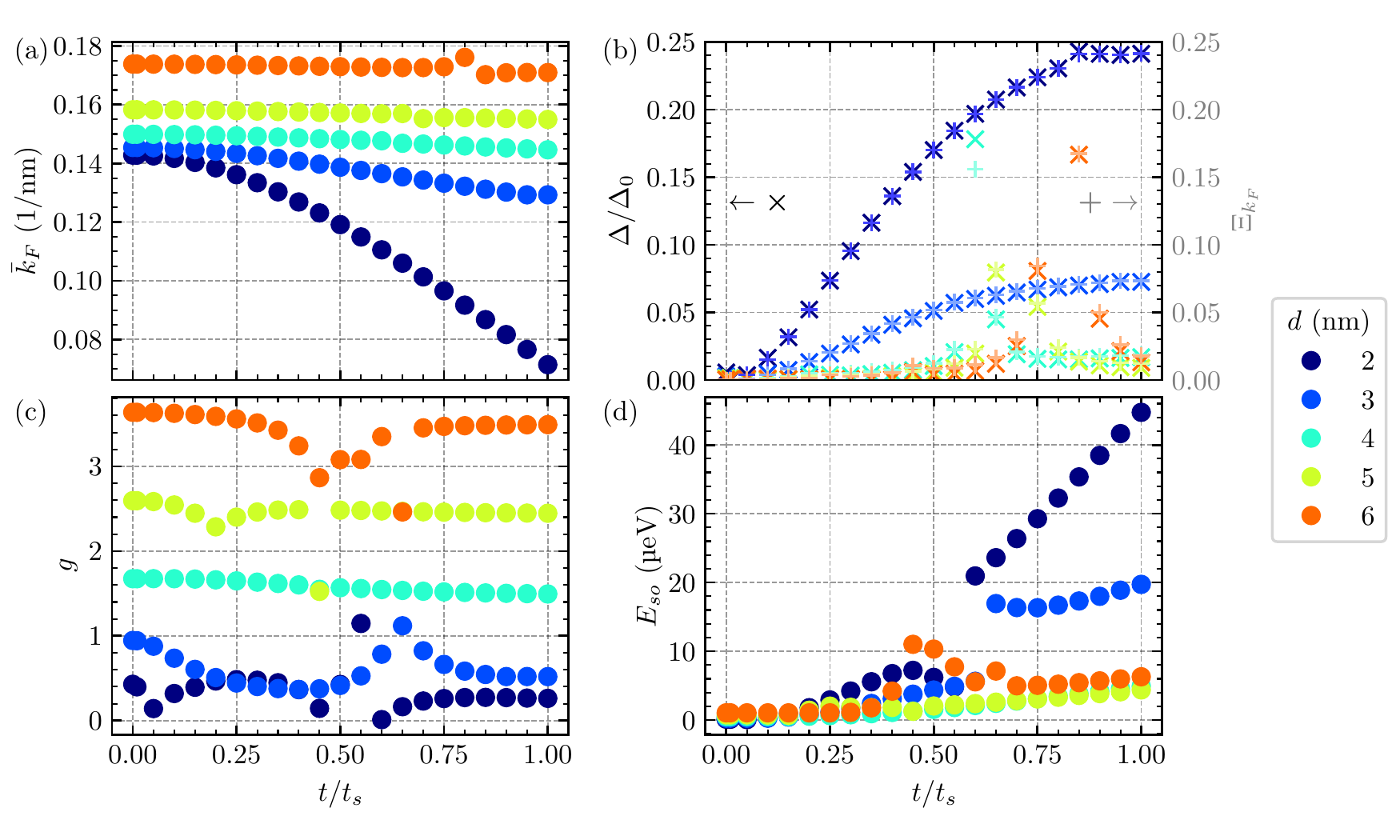}
	\caption{(a) Average Fermi wave vector $\bar{k}_F$, (b) proximity-induced superconducting gap (crosses, left axis) and the weight $\Xi_{k_F}$ of the NW state in the SC at $k_F$ (pluses, right axis), (c) NW $g$ factor, and (d) spin-orbit energy as a function of the coupling $t$ between SC and NW, and for various heights $d$ of the Ge NW. The chemical potential of the SC is chosen such that $\bar{k}_F$ decreases with increasing coupling. The effect is strongest for a thin NW. As expected, the weight $\Xi_{k_F}$ of the NW state in the SC shows the same functional behavior as the ratio of the induced superconducting gap and the parent SC pairing potential, $\Delta/\Delta_0$. The gap increases with the coupling and is largest for a thin NW. Only for a thin NW is the NW wave function main weight close to the SC-NW interface and the proximity effect can be sizable. The $g$ factor is smallest for a thin NW ($d=2$~nm), where the 2D physics dominates. It decreases slightly with increasing coupling. Due to the coupling between NW and SC, the NW wave function gains more weight at the interface breaking the inversion symmetry, which is similar to the response of the wave function to an external electric field. This symmetry breaking gives rise to a weak SOI that increases with the coupling $t$. At certain values of $t$, we observe deviations of the induced gap, $g$ factor, and spin-orbit energy from the general $t$ dependence. This is always the case close to a resonance, as discussed in Sec.~\ref{sec:SCchempot}. However, the $g$ factor and the spin-orbit energy are more sensitive to resonances at $k=0$, while the induced gap is more sensitive to resonances at $k=k_F$. Unless stated otherwise, the parameters are the same as in  Fig.~\ref{fig:dispRelmuSC}. \label{fig:NWthickness}}
\end{figure*}

In this section, we study how certain parameters, namely, the average Fermi wave vector $\bar{k}_F$, the induced superconducting gap $\Delta$, the $g$ factor, and the spin-orbit energy $E_{so}$, of the proximitized NW depend on the thickness $d$ of the NW. Throughout, we keep the dimensions of the SC to $L_x = L_y = \SI{10}{\nano\meter}$ and the width of the NW $L_x = \SI{10}{\nano\meter}$ (see Fig.~\ref{fig:schematic}).  We show the results in Fig.~\ref{fig:NWthickness}. As discussed in Sec.~\ref{sec:SCchempot}, we set the chemical potentials of the NW and the SC such that the average Fermi wave vector shrinks with increasing $t$. This is reflected by the results shown in Fig.~\ref{fig:NWthickness}(a) for all values of considered NW thicknesses $d$. 
 
In Fig.~\ref{fig:NWthickness}(b), we show that with increasing coupling $t$, a superconducting gap $\Delta$ is induced in the NW and this gap increases with $t$ until it reaches a maximum that depends on the thickness of the NW. The induced gap is largest for $d=\SI{2}{\nano\meter}$, where it reaches $\Delta = 0.24\, \Delta_0$. When $d\geq \SI{4}{\nano\meter}$, the gap is rather small because the NW wave function is localized in the center of the NW cross section far away from the interface coupled to the SC. Note that in Sec.~\ref{sec:elecField}, we will discuss a way to also reach a sizable gap in thicker NWs. As we expect, the size of the gap (crosses) shows the same functional behavior as the weight $\Xi_{k_F}$ of the NW state in the SC at $k_F$ (pluses) [see inset in Fig.~\ref{fig:dispRelmuSC}(a)]. For $d= \SI{4}{\nano\meter}, \SI{5}{\nano\meter},$ and $\SI{6}{\nano\meter}$, the induced gap has peaks for certain values of the coupling $t$, which can be traced back to the strong hybridization of the NW state with a SC state it is coupled to at a resonance, as discussed in Sec.~\ref{sec:SCchempot}. However, for the induced gap, resonances at $k=k_F$ are more relevant than resonances at $k=0$. By further increasing $t$, the SC and the NW states move away from each other in energy and the $\Delta$ profile returns to the general behavior. The same interpretation holds for the peaks and dips of the $g$ factor [see Fig.~\ref{fig:NWthickness}(c)] and the spin-orbit energy $E_{so}$ [see Fig.~\ref{fig:NWthickness}(d)]. The states in the Ge NW are mixed HH-LH states and thus the superconductivity has support from both types of holes.
 
 The Ge hole NW $g$ factor at $k=0$ depends only slightly on $t$. In general, it decreases as the coupling becomes stronger. The $g$ factor is largest ($g \approx 3.6$) for a thick NW ($d=\SI{6}{\nano\meter}$), where the NW is governed by strong HH-LH mixing~\cite{Kloeffel2018,Adelsberger2022a}, whereas for $d= \SI{2}{\nano\meter}$, the lowest-energy eigenstates are almost purely of an HH nature, resulting in a small in-plane $g$ factor ($g<1$)~\cite{Scappucci2021}. 
 
 Without coupling between the NW and the SC, the spin-orbit energy is zero in the NW since there is no electric field~\cite{Kloeffel2011}. As the coupling increases, the NW wave function is pushed closer towards the NW-SC interface and thereby gets squeezed. This breaks the inversion symmetry similar to an external electric field that pushes the wave function towards the SC. Thus, a finite spin-orbit energy $E_{so}$ develops, as shown in Fig.~\ref{fig:NWthickness}(d). The spin-orbit energy is larger for thinner NWs since there the coupling to the SC has the strongest effect due to the proximity of the wave functions of states in the NW to the SC. The spin-orbit energy is determined by the energy difference between the maximum of the negatively curved Rashba band and the spin-orbit crossing point at $k=0$ of the hole NW.
 
\section{Electric field \label{sec:elecField}}

As discussed in Sec.~\ref{sec:NWthick}, only in very thin NWs is it possible to induce a sizable superconducting gap since the wave function in the NW needs to be close to the NW-SC interface. However, there are several reasons for using thicker NWs. For instance, the $g$ factor increases with the thickness $d$, as shown in Fig.~\ref{fig:NWthickness}(c). Also, with thicker NWs, it is possible to achieve the side length ratio of the NW that maximizes the SOI for a certain value of the electric field~\cite{Adelsberger2022}.

 \begin{figure*}[]
	\includegraphics{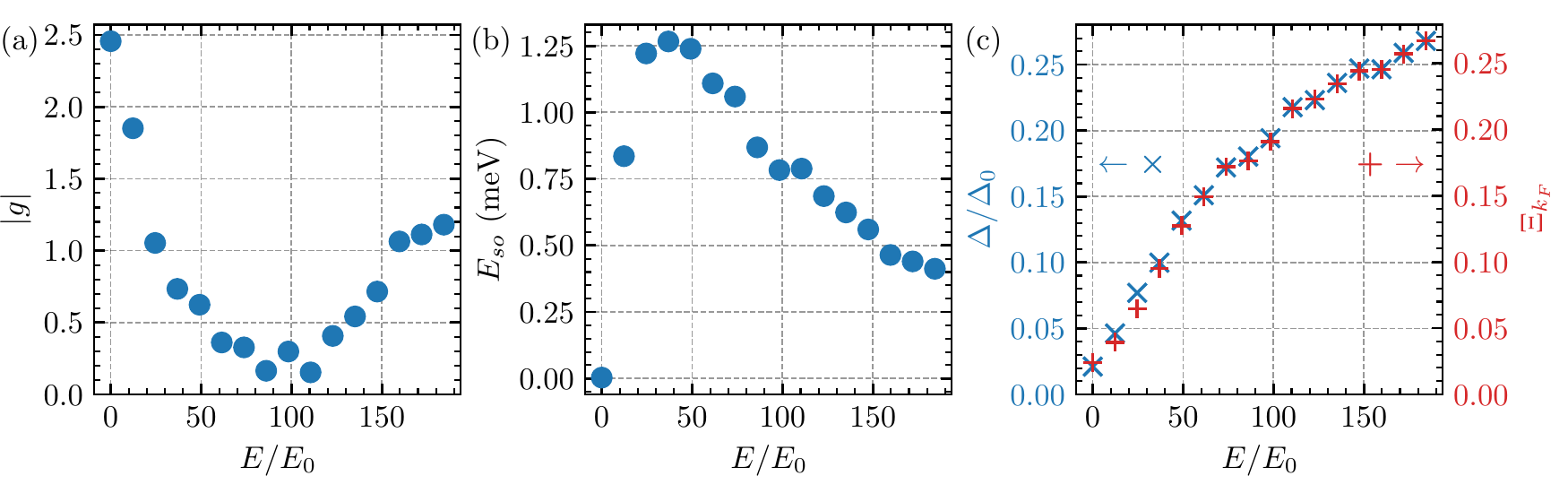}
	\caption{Dependence on the electric field $E$: (a) $g$ factor, (b) spin-orbit energy $E_{so}$, and (c) proximity-induced gap $\Delta$ (blue crosses), as well as the weight $\Xi_{k_F}$ of the NW state in the SC at $k_F$ (red pluses) as a function of $E$ applied in the $y$ direction for $t=0.8\, t_s$. The electric field is in units of $E_0 = \frac{\hbar^2 \gamma_1}{2 m e d^3} \approx \SI{4}{\volt\per\micro\meter}$ for $d=\SI{5}{\nano\meter}$. The electric field has basically three effects: It squeezes the wave function in the NW, which reduces the $g$ factor [see (a)], induces a strong SOI [see (b)], and pushes the wave function within the NW cross section close to the interface with the SC which enhances the leakage of the wave function into the SC and thereby the proximity-induced gap [see (c)]. Thus, it is possible to have strong SOI and a large proximity-induced gap simultaneously. Finding the best parameters for a sizable $g$ factor, SOI, and superconducting proximity gap requires some optimization. Panel (a)  shows the absolute value of the $g$ factor. At $E\approx80\, E_0$, the $g$ factor changes sign. At strong electric field ($E\gtrsim150\, E_0$),  it is dominated by the SC $g$ factor since the wave function at $k=0$ has a weight of almost \SI{70}{\percent} in the SC. The position of the maximum of the spin-orbit energy is approximately reached at $\abs{e U E/\Delta_{sb}} \sim 1$~\cite{Kloeffel2018}. If not stated otherwise, the parameters are the same as in  Fig.~\ref{fig:dispRelmuSC} and $d=\SI{5}{\nano\meter}$. \label{fig:NW_E}}
\end{figure*}

Apart from reducing thickness, there is another way to move the wave function closer to the interface, namely, an external electric field in the $y$ direction. In Fig.~\ref{fig:NW_E}, we set the NW-SC coupling to $t= 0.8\, t_s$, as well as $d=\SI{5}{\nano\meter}$, and plot the $g$ factor, SOI, and proximity-induced gap as a function of the external electric field $E$. We give the electric field in units of $E_0 = \frac{\hbar^2 \gamma_1}{2 m e d^3} \approx \SI{4}{\volt\per\micro\meter}$ for $d=\SI{5}{\nano\meter}$. Note that $E/E_0 = d^3/l_E^3$  with the electric length $l_E = (\hbar^2 \gamma_1/2meE)^{1/3}$, where $m/\gamma_1$ is the average HH-LH mass. Typically, the mass of the hole NW ground state converges to the average HH-LH mass for strong electric field~\cite{Adelsberger2022}. Since the external electric field shifts the NW bands in energy and we want to focus on the lowest-energy NW state, we compensate for this effect by adjusting the chemical potential $\mu_w$. As a function of the electric field, the $g$ factor is first reduced until it changes sign at  $E\approx80\, E_0$ [see Fig.~\ref{fig:NW_E}(a)]. The small increase for the $g$ factor at $E\approx100\, E_0$, followed by another dip, is associated with a resonance (see discussion in Sec.~\ref{sec:SCchempot}). Since we show the absolute value, the $g$~factor increases for stronger electric fields. It reaches a value close to one, which is set by the SC $g$ factor because at strong electric field ($E\gtrsim150\, E_0$), the wave function  at $k=0$ has a weight of almost \SI{70}{\percent} in the SC.

The spin-orbit energy, on the other hand, is small in the absence of an electric field [see Fig.~\ref{fig:NWthickness}(d)] and  reaches a maximum at $E\approx37\,E_0$, after which it is gradually reduced with further increasing the electric field [see Fig.~\ref{fig:NW_E}(b)]. This is the typical behavior of the SOI in hole NWs, which is referred to as the direct Rashba SOI~\cite{Kloeffel2011,Kloeffel2018,Adelsberger2022,Adelsberger2022a}. This very strong type of SOI originates in the HH-LH mixing in 1D hole systems in combination with the breaking of inversion symmetry. The position of the maximal spin-orbit energy is approximately reached when $\abs{e U E/\Delta_{sb}}$ becomes of the order of one~\cite{Kloeffel2018}, where $\Delta_{sb}$ is the subband gap in the NW and $U=0.15 d/2$.
Interestingly, the proximity-induced gap increases with the electric field and reaches a value above $\Delta= 0.25\, \Delta_0$, which is comparable to the situation of the flat NW ($d=\SI{2}{\nano\meter}$) in Fig.~\ref{fig:NWthickness}(b). Again, the induced gap shows the same dependence on $E$ as the weight of the NW wave function in the SC, $\Xi_{k_F}$ [see Fig.~\ref{fig:NW_E}(c)].

The SOI behaves as expected from former studies~\cite{Adelsberger2022,Kloeffel2018}. However, in a standard Rashba NW, the spin-orbit energy decreases as the proximity-induced gap increases~\cite{Reeg2018}. Here, we find that this is not necessarily true for holes in Ge since an appropriate external electric field can cause strong SOI and a sizable superconducting gap in a Ge NW at the same time. Also relevant for the search of MBSs, where in addition to strong SOI and a proximity-induced gap a large Zeeman gap is required, it is important to avoid regimes of strongly reduced $g$ factor. As shown in Fig.~\ref{fig:NW_E}, it is possible to achieve a large SOI and reasonably large induced gap for realistic field strengths, while the $g$ factor remains relatively small; however, it can also be optimized by adjusting the field strength. Furthermore, the coexistence of a large gap and strong SOI by itself is promising for proposals that achieve MBSs in Ge without any requirement for a large Zeeman energy or with reduced requirements on the Zeeman energy \cite{Luethi2023,Luethi2023a,Lesser2022}.

\begin{figure*}
	\includegraphics{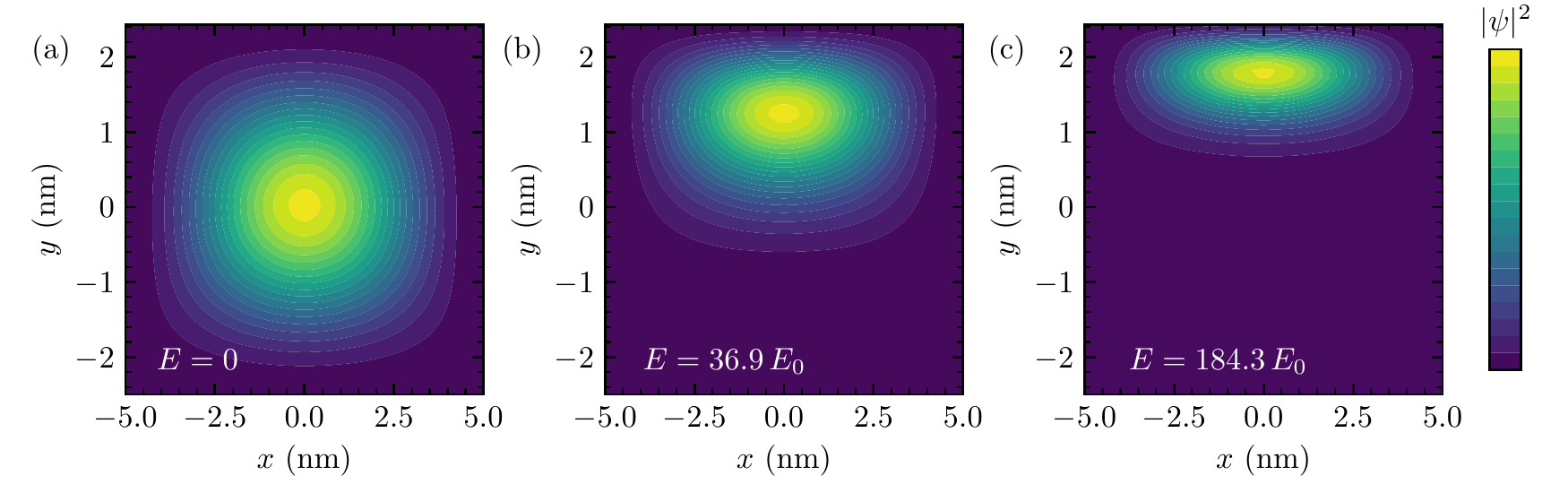}
	\caption{Wave function in the NW for NW-SC coupling $t = 0.8\, t_s$ at (a) $E=0$, (b) $E=36.9\, E_0$, and (c) $E=184.3\, E_0$ at $k=k_F$. The superconductor is located on top of the NW. Without the electric field, the wave function is located in the center of the NW and the coupling to the SC is not sufficient to shift it towards the SC. For $E=36.9\, E_0$, where we observe the maximum SOI in Fig.~\ref{fig:NW_E}(b), the wave function is squeezed and pushed towards the NW-SC interface. This explains the gap opening and reduction of the $g$ factor shown in Fig.~\ref{fig:NW_E}. Further increasing the electric field to $E=184.3\, E_0$ further pushes the wave function to the top and enhances the squeezing, resulting in a larger proximity-induced gap [see Fig.~\ref{fig:NW_E}(c)]. If not stated differently, the parameters are the same as in  Fig.~\ref{fig:dispRelmuSC} and $d=\SI{5}{\nano\meter}$.\label{fig:NW_WF}}
\end{figure*}

In  order to get a better understanding for the observed behavior in Fig.~\ref{fig:NW_E}, we plot in Fig.~\ref{fig:NW_WF} the wave function of the NW state at $k = k_F$ (by $k_F$, we denote the Fermi wavevector at which the proximity-induced gap is smallest) for different values of the electric field. We choose $E=0$ [see Fig.~\ref{fig:NW_WF}(a)], the electric field where the spin-orbit energy is maximal 
$E\approx37\,E_0$ [see Fig.~\ref{fig:NW_WF}(b)], and the maximum considered electric field $E\approx184\,E_0$ [see Fig.~\ref{fig:NW_WF}(c)]. In the absence of an external electric field for $d=\SI{5}{\nano\meter}$, the coupling to the SC is not sufficient to push the NW wave function towards the interface with the SC. Thus, the wave function mostly remains in the center of the NW cross section, resulting in a rather weak SOI [see Fig.~\ref{fig:NW_E}(b)] and small proximity-induced gap [see Fig.~\ref{fig:NW_E}(c)]. At $E\approx37\,E_0$, the wave function is squeezed in the $y$ direction, which causes a drop in the $g$ factor. At the same time, the wave function is pushed towards the NW-SC interface, breaking the symmetry. The proximity of the wave function to the SC allows for a stronger leakage of the NW wave function into the SC, resulting in an enhanced proximity-induced superconducting gap in the NW. At $E\approx184\,E_0$, the wave function is further pushed towards the SC and the coupling, along with the induced gap, is increased. Also, the wave function is squeezed very strongly, which is why we would expect an even smaller $g$ factor. However, the NW state now hybridizes strongly with the SC states, resulting in a $g$ factor of the order of one.

For a Ge-based NW setup that is promising for the formation of Majorana bound states, we identify the following electric field regimes as optimal: Between $E\approx 25\, E_0$ and $E\approx 60\, E_0$, the $g$ factor is between $0.3$ and $1.1$. At the same time, the spin-orbit energy reaches its maximum within this electric field range and the gap starts to open with values between $\Delta=0.08\,\Delta_0$ and $\Delta=0.15\,\Delta_0$. The regime $E\gtrapprox 120 E_0$ is also promising. We find decent values of the spin-orbit energy $E_{so}$ simultaneously with a gap around $\Delta= 0.25\,\Delta_0$, and with $g$ factors larger than one, largely due to the finite $g$ factor of the SC.

\begin{figure*}[]
	\includegraphics{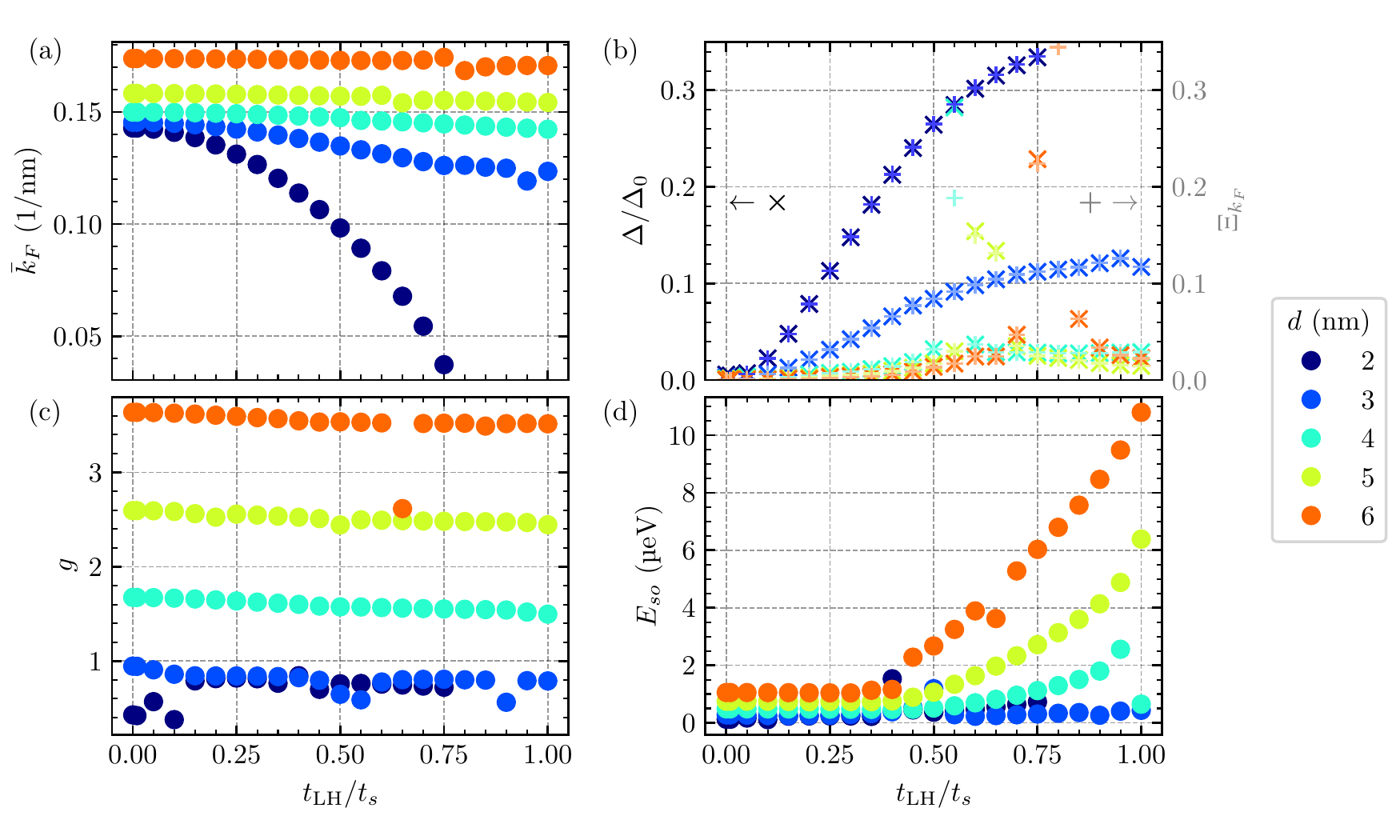}
	\caption{Same figure as Fig.~\ref{fig:NWthickness}, but the SC is coupled only to the LHs in the NW ($t_\mathrm{HH}=0$). Qualitatively, there is no difference from Fig.~\ref{fig:NWthickness}. The proximity-induced gap is larger when the SC couples only to the LHs [see (b)]. The spin-orbit energy in (d) is strongly reduced without coupling to the HHs in the thin NWs ($d=\SI{2}{\nano\meter}$ and $d=\SI{3}{\nano\meter}$). In (a), the Fermi wave vector is not well defined for $t_\mathrm{LH}>0.75\, t_s$, similarly to the cyan line in Fig.~\ref{fig:dispRelmuSC}(c).\label{fig:NWthickness_noHH}}
\end{figure*}

For the topological phase transition, if the chemical potential is tuned to the crossing point of the spin-orbit split subbands in the NW, it is required that the Zeeman gap becomes larger than the proximity-induced superconducting gap $\Delta_Z = \frac{1}{2} g \mu_B B> \Delta$~\cite{Lutchyn2010,Oreg2010,Maier2014}. Taking as an example the values we get at $E= 160\, E_0$, where $E_{so} = \SI{460}{\micro\electronvolt}$, $g = 1.06$, and $\Delta=0.25\,\Delta_0=\SI{50}{\micro\electronvolt}$, a magnetic field of $B \sim \SI{2}{\tesla}$ is necessary to fulfill the topological condition. Thus, despite the very small $g$ factor, it is within the realms of possibility to reach the topological phase in a Ge NW system since the critical field parallel to, e.g., thin Al films can be as large as $B_c=\SI{5}{\tesla}$~\cite{Meservey1971}. Note that this is just a rough estimate since an externally applied magnetic field influences the effective $g$ factor in the Ge NW~\cite{Adelsberger2022a} and the induced superconducting gap will be suppressed. However, a detailed analysis taking these effects into account also predicts the topological phase to be possibly within reach~\cite{Laubscher2023}.

We mention here that for a more symmetric NW cross section as in a cylindrical NW, the state is quasidegenerate, which spoils the potential formation of MBSs. However, due to static strain coming from a shell around the NW or from the contact to the SC, which is neglected in this work, a substantial subband gap emerges, lifting this quasidegeneracy~\cite{Kloeffel2011,Kloeffel2014,Adelsberger2022}.

In addition to the external electric field considered here, the band bending at the interface between the SC and semiconductor is expected to create an electrostatic potential in a realistic device. We neglect these effects in this work because the electrostatics would only renormalize the electric field strength required to localize holes close to the interface. Therefore, the qualitative effect of simultaneously enhanced proximity-induced superconductivity and SOI  can be expected to remain unaltered.

\begin{figure*}[]
	\includegraphics{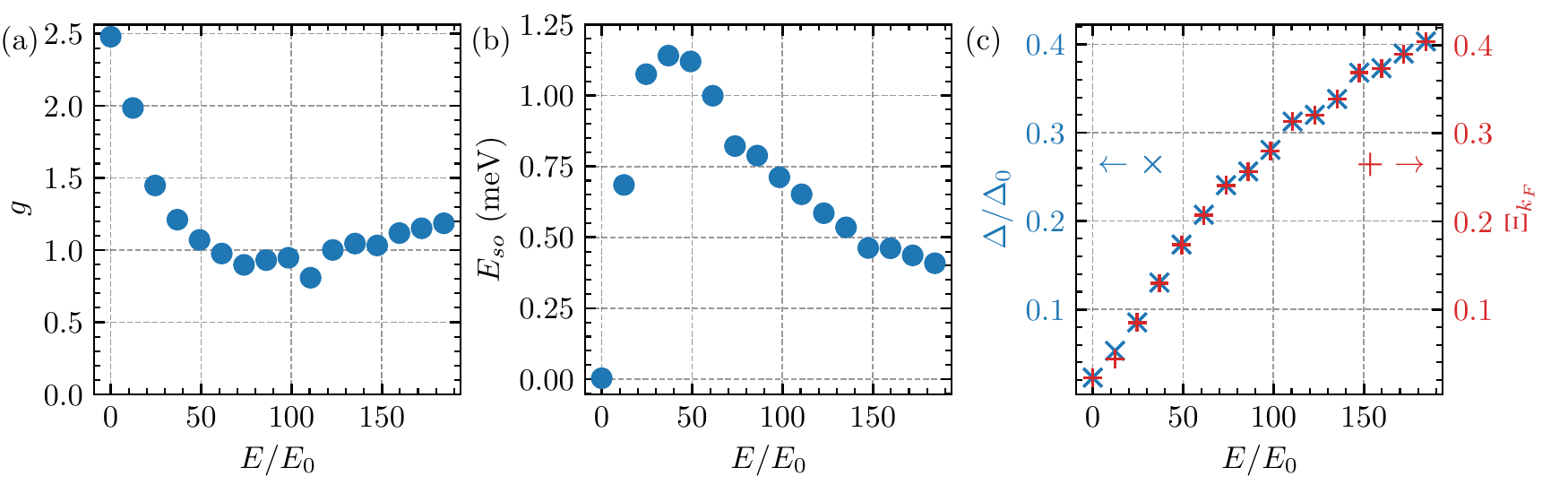}
	\caption{Same figure as Fig.~\ref{fig:NW_E}, but the SC is coupled only to the LHs in the NW ($t_\mathrm{HH}=0$, $t_\mathrm{LH} = 0.8\, t_s$). The qualitative behavior is similar to Fig.~\ref{fig:NW_E}. (a) The $g$ factor never drops to a value as low as with coupling also to HHs. For $E>50\, E_0$, the $g$ factor stays at a value of around one. (b) The spin-orbit energy is almost identical to the one in Fig.~\ref{fig:NW_E}, but the maximum is slightly lower at $E_{so} = \SI{1.1}{\milli\electronvolt}$ instead of $E_{so} = \SI{1.3}{\milli\electronvolt}$. (c) The proximity-induced gap reaches $\Delta = 0.40\, \Delta_0$ at $E\approx184\,E_0$, which is significantly larger than the maximum of $\Delta = 0.27\, \Delta_0$ in Fig.~\ref{fig:NW_E}. \label{fig:NW_EnoHH}}
\end{figure*}

\section{Coupling only to light holes in the nanowire \label{sec:noHH}}

In this section, we analyze the case where $t_\mathrm{LH}\neq 0 $ and $t_\mathrm{HH}=0$ [see Eq.~\eqref{eqn:NWSCcoup}] as an extreme case of $t_\mathrm{LH}\neq t_\mathrm{HH}$. In Fig.~\ref{fig:NWthickness_noHH}, we show the same plots as in Fig.~\ref{fig:NWthickness}, but with $t_\mathrm{HH}=0$. In general, we observe the same qualitative behavior with and without coupling to the HHs. Interestingly, the proximity-induced gap $\Delta$ can be larger for $t_\mathrm{HH} =0$ than for  $t_\mathrm{HH} \neq 0$ [see Fig.~\ref{fig:NWthickness_noHH}(b)]. However, still the induced gap is rather small for $d \geq \SI{4}{\nano\meter}$.  Another interesting difference from Fig.~\ref{fig:NWthickness} can be seen in Fig.~\ref{fig:NWthickness_noHH}(d). The spin-orbit energy, which is induced by coupling to the SC, is much smaller for $d=\SI{2}{\nano\meter}$ and $d=\SI{3}{\nano\meter}$ in the absence of the coupling to the HHs. 

Also in the case without coupling to the HHs, we analyze the effect of an electric field that pushes the NW wave function towards the interface with the SC. In Fig.~\ref{fig:NW_EnoHH}, we show the same data as in Fig.~\ref{fig:NW_E}, but without coupling to the HHs. Again, we find  a similar qualitative behavior as before. For $t_\mathrm{HH}=0$, the $g$ factor minimum is $0.8$ and, for $E\gtrsim 50\, E_0$, it stays close to a value of $g\sim 1$ [see Fig.~\ref{fig:NW_EnoHH}(a)]. The $g$ factor does not drop to zero since the HHs remain uncoupled from the SC and the $g$ factor does not change sign. For $E=0$, the $g$ factor is $g = 2.5$ regardless of the SC coupling to HHs. For $t_\mathrm{HH}=0$, the spin-orbit energy $E_{so}$ reaches only a slightly smaller value than the previous maximum, which again occurs close to $E = 37\, E_0$  [see Fig.~\ref{fig:NW_EnoHH}(b)]. For $t_\mathrm{HH}=0.8\, t_s$, the maximum spin-orbit energy is $E_{so} = \SI{1.3}{\milli\electronvolt}$, while it is  $E_{so} = \SI{1.1}{\milli\electronvolt}$ for $t_\mathrm{HH}=0$. The proximity-induced gap, on the other hand, is larger for $t_\mathrm{HH}=0$ [see Fig.~\ref{fig:NW_EnoHH}(c)]. At $E = 184\, E_0$, the gap is $\Delta = 0.40\, \Delta_0$ instead of $\Delta = 0.27\, \Delta_0$ for the previous case of $t_\mathrm{HH}=0.8\, t_s$. 

There are no crucial qualitative differences between the  situations with $t_\mathrm{HH} = 0$ and $t_\mathrm{HH} \neq 0$ that would contradict our main message of simultaneously large spin-orbit energy and proximity-induced gap. In a realistic experiment, we expect $t_\mathrm{HH} \neq t_\mathrm{LH}$ and  $t_\mathrm{HH} \neq 0$, which should result in a situation between the two that was discussed in this paper. Again, in such a realistic situation, we will likely find a regime where the $g$ factor, the SOI, and the proximity-induced gap are just large enough to achieve the topological phase transition. 

\section{Conclusion \label{sec:conclusion}}
We have numerically investigated the coupling between a Ge NW and a SC that were both modeled as 3D systems. We showed that the average Fermi wave vector depends on the coupling to the SC and can increase or decrease with increasing coupling depending on the chemical potential of the SC with respect to the chemical potential of the NW. 

We found that effects of the coupling between the NW and the SC strongly depend on the thickness of the NW. When no electric field is applied, only for thin NWs does the coupling result in a sizable proximity-induced gap in the NW. This observation can be explained by the distance of the NW wave function to the SC. Only if the wave function is close to the SC does it strongly couple. We showed that the $g$ factor is largest for thicker NWs where, in the absence of an electric field, the proximity-induced gap remains negligibly small, even for large hopping between SC and NW sites at the interface. Since the induced gap depends on the distance of the NW wave function to the SC, the gap can be increased by pushing the wave function closer towards the SC by an external electric field. We demonstrated that it is possible to achieve a large spin-orbit energy simultaneously with a sizable proximity-induced gap for a certain range of the electric field.

There, however, exists an electric field regime where the $g$ factor drops to almost zero, which spoils the applicability of the Ge hole NW coupled to a SC for the formation of Majorana bound states. Still, some optimization allows for a regime where the $g$ factor, spin-orbit energy, and proximity-induced gap are just large enough to achieve a topological phase transition. A rough estimate tells us that in the most optimal scenario, the topological phase can be reached with a magnetic field $B\sim\SI{2}{\tesla}$, which is below the critical field of thin Al films. A different type of coupling where the SC does not couple equally to HHs and LHs does not change the qualitative results. Coupling only to LHs even can have a positive effect on the $g$ factor of the proximitized NW.

Our results indicate that there is a unique phenomenology of metallization effects in Ge. Most importantly, it is possible to find scenarios where the spin-orbit energy and the induced superconducting gap are sizable. This is encouraging for  many potential applications of hybrid Ge-SC devices in quantum information processing, e.g., Andreev spin qubits. It is also promising for the realization of topological superconductivity in Ge-based systems, especially for protocols where no Zeeman energy is required~\cite{Luethi2023a,Lesser2022} since then no trade-off is necessary in order to optimize the $g$ factor.

\begin{acknowledgments}
We thank Stefano Bosco and Dominik Zumb\"uhl for useful discussions. This work is supported by the Swiss National Science Foundation (SNSF) and NCCR SPIN (Grant No. 51NF40-180604). This project has received funding from the European Union’s Horizon 2020 research and innovation programme under Grant Agreement No 862046. H.F.L. acknowledges support from the Georg H. Endress Foundation.
\end{acknowledgments}

\bibliography{Literature}
	
\end{document}